\documentclass[a4paper,10pt]{article}
\usepackage{amsfonts,amsthm,amsmath,amssymb,graphicx}

\newcommand{\be}{\begin{equation}}
\newcommand{\bea}{\begin{eqnarray}}
\newcommand{\ee}{\end{equation}}
\newcommand{\eea}{\end{eqnarray}}

\begin{document}

\makeatletter
\@addtoreset{equation}{section}
\makeatother
\renewcommand{\theequation}{\thesection.\arabic{equation}}

\rightline{WITS-CTP-145}
\vspace{1.8truecm}

\vspace{15pt}


{\large{
\centerline{   \bf Evolution of Quark Masses and Flavour Mixings in the 2UED}
\centerline {\bf  }
}}

\vskip.9cm

\thispagestyle{empty} \centerline{
     {\bf Ammar. Abdalgabar${}^{} $\footnote{ {\tt ammar.abdalgabar@students.wits.ac.za}} and A. S. Cornell${}^{} $\footnote{ {\tt alan.cornell@wits.ac.za}}}
                                                    }

\vspace{.8cm}
\centerline{{\it ${}$ National Institute for Theoretical Physics;}}
\centerline{{\it School of Physics,University of the Witwatersrand }}
\centerline{{\it Wits 2050, South Africa } }

\vspace{1.4truecm}

\thispagestyle{empty}

\centerline{\bf Abstract}\label{abst}
\vskip.4cm
The evolution equations of the Yukawa couplings and quark mixings are performed for the one-loop renormalisation group equations in six-dimensional models compactified in different possible ways to yield standard four space-time dimensions. Different possibilities for the matter fields are discussed, that is where they are in the bulk or localised to the brane. These two possibilities give rise to quite similar behaviours when studying the evolution of the Yukawa couplings and  mass ratios. We find that for both scenarios, valid up to the unification scale, significant corrections are observed.  

\vspace{1cm}
{\large {\bf Keywords}}:  Fermion masses, extra Dimension. Beyond Standard Model


\pagenumbering{arabic}


\setcounter{footnote}{0}

\linespread{1.1}
\parskip 4pt

{}~
{}~


\section{Introduction}
\par A theory of fermion masses and the associated  mixing angles is unexplained in the Standard Model (SM) providing an interesting puzzle and a likely window to physics beyond the SM. In the SM one of the main issues is to understand the origin of quark and lepton masses, or the apparent hierarchy of family masses and quark mixing angles. Perhaps if we understood this we would also know the origins of CP violation. A clear feature of the fermion mass spectrum is \cite{Falcone:2001ep,Liu:2009vh}

\begin{equation}
m_u \ll m_c \ll m_t \; , \; m_d \ll m_s \ll m_b \; , \; m_e \ll m_{\mu} \ll m_{\tau}.
\end{equation}

\par Apart from the discovery of the Higgs boson at the Large Hadron Collider (LHC), another important goal of the LHC is to explore the new physics that may be present at the TeV scale. Among these models those with extra spatial dimensions offer many possibilities for model building and TeV scale physics scenarios which can be constrained or explored. As such, there have been many efforts to understand the fermion mass hierarchies and their mixings by utilizing the Renormalization Group Equations (RGEs) especially for Universal Extra Dimension (UED) models and their possible extensions (see Refs.\cite{Cornell:2010sz,Cornell:2011fw} and references therein).

 \par UED models at the TeV scale are discussed in various configurations, the simplest being the case of one flat extra dimension compactified on $S^1/Z_2$, which has been widely studied and constrained for more than a decade \cite{ Appelquist:2000nn}. Electroweak precision measurements \cite{Appelquist:2002wb} combined with the LHC Higgs bounds impose a lower bound of $R^{-1} \geq 700$ GeV on the compactification scale \cite{Aad:2012an} . On the other hand the dark matter relic density observed by WMAP \cite{Komatsu:2010fb} sets an upper bound on the compactification sale of $ 1.3$ TeV $\leq R^{-1} \leq 1.5$ TeV. In these UED models each SM field is accompanied by a tower of massive states, the Kaluza-Klein (KK) particles. An extension of this scenario is to consider a type of model with two extra dimensions. This extension is non-trivial and brings further insight to extra-dimensional scenarios. It is theoretically motivated by specific requirements, such as, they provide a dark matter candidate, suppress the proton decay rate, as well as anomaly cancellations from the number of fermion generations being a multiple of three \cite{Dobrescu:2001ae}. Different models with two extra dimensions have been proposed, such as  $T^2/Z_2$ \cite{Appelquist:2000nn}, the chiral square $T^2/Z_4$ \cite{Dobrescu:2004zi}, $T^2/(Z_2 \times Z'_2)$ \cite{Mohapatra:2002ug}, $S^2/Z_2$ \cite{Maru:2009wu}, the flat real projective plane $RP^2$ \cite{Cacciapaglia:2009pa}, the real projective plane starting from the sphere \cite{Dohi:2010vc}.
For simplicity, in this paper we assume that the two extra space-like dimensions have the same size, that is $R_5 = R_6 =R$.  However, this simpler case provides the opportunity to compute in detail the RGEs and study the evolution of mass ratios, the renormalisation invariance $R_{13}$ and $R_{23}$, and $\sin \beta$.

\par The four-dimensional chiral zero modes of the SM fermion are obtained by imposing a discrete $Z_2$ symmetry, this eliminates one 4-dimensional (4D) degree of freedom  and allows us to have a 4D chiral fermion \cite{Dobrescu:2004zi}. However, this can also be obtained directly from the properties of the orbifold, as in Ref.\cite{Cacciapaglia:2009pa}. Higher massive modes are then vector-like fermions. Each of the gauge fields have six components and decompose into towers of 4D spin-1 fields and two towers of real scalars belonging to the adjoint representation \cite{Abdalgabar:2013oja}. 

\par The one-loop correction to the gauge couplings are given by
\begin{equation}
16 \pi^2 \frac{d g_i}{d t}= b^{SM}_i g^3_i+\pi \left( S(t)^2-1 \right) b^{6D}_i g^3_i \;,
\label{gauge2UED}
\end{equation}
where $t = \ln (\frac{\mu}{M_Z})$, $S(t) = {e^t}{M_Z}R$, or $S(\mu)=\mu R=\frac{\mu}{M_{KK}}$ for $M_Z < \mu < \Lambda$ ($\Lambda$ is the cut-off scale, where we have set $M_Z$ as the renormalisation point). More details about the calculation of the $S^2(t)$ factor can be found in Refs.\cite{Abdalgabar:2013oja,Abdalgabar:2013xsa}. The numerical coefficients appearing in Eq.(\ref{gauge2UED}) are given by:
\begin{equation}
b^{SM}_i=  \left[ \frac{41}{10}, -\frac{19}{6}, -7\right]\;,
\end{equation}
and 
\begin{equation}
b^{6D}_i= \left[\frac{1}{10}, -\frac{13}{2}, -10\right]+\left[\frac{8}{3}, \frac{8}{3}, \frac{8}{3}\right]\eta\;,
\end{equation}
\noindent $\eta$ being the number of generations of fermions propagating in the bulk. Therefore, in the two cases we shall consider: that of all fields propagating in the bulk, $\eta =3$; and for all matter fields localized to the brane $\eta =0$.

\begin{figure}[h!]
\begin{center}
\includegraphics[width=7cm,angle=0]{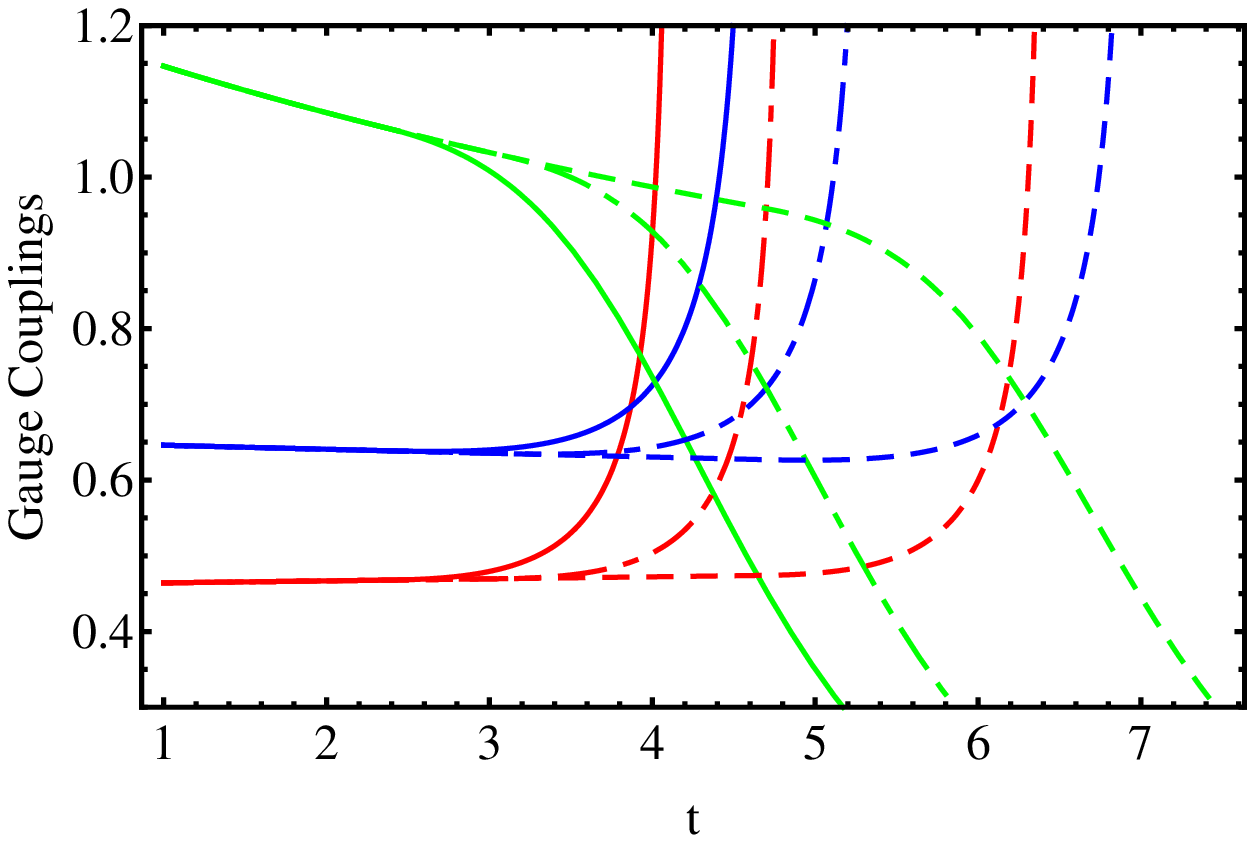}\qquad 
\includegraphics[width=7cm,angle=0]{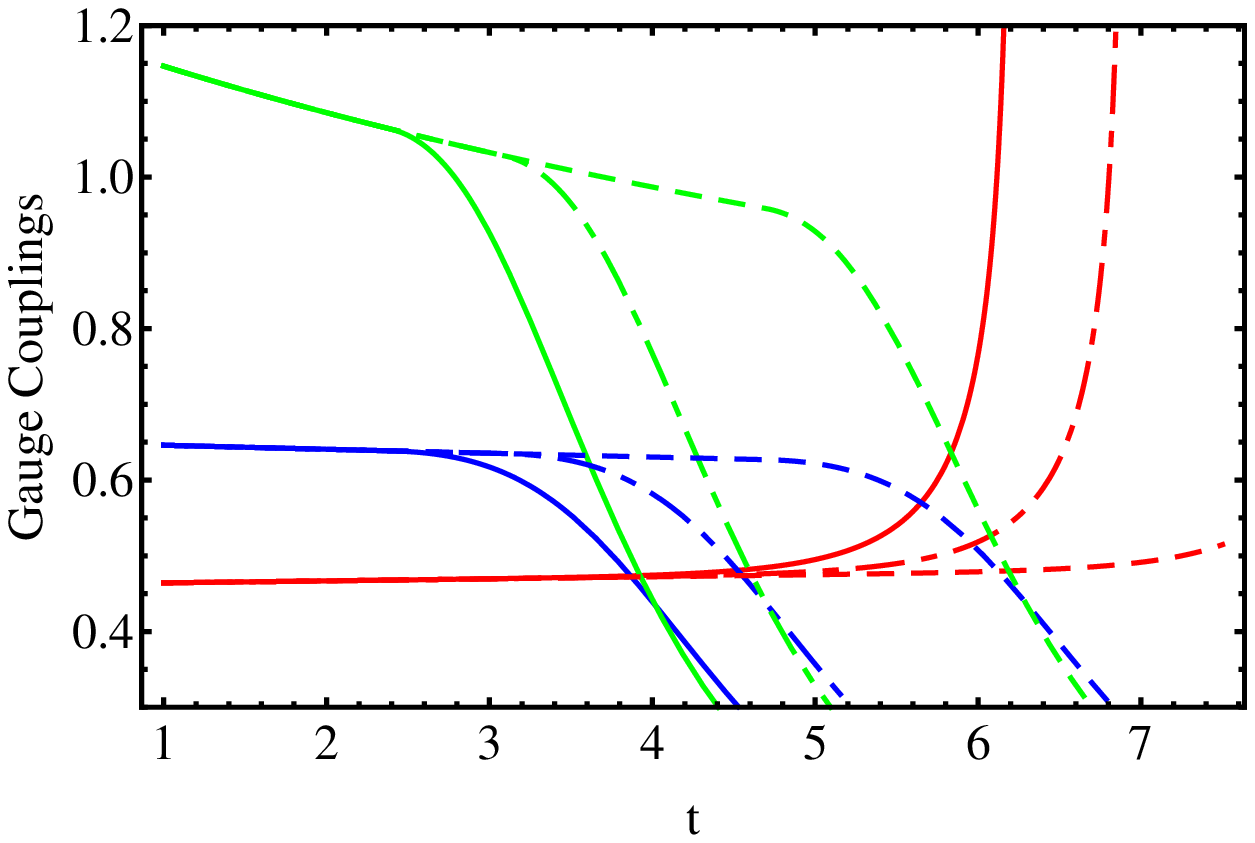}
\end{center}
\caption{\label{fig1} \it (Colour online) Gauge couplings {$g_1$} (red), {$g_2$} (blue), {$g_3$} (green) with: in the left panel, all matter fields in the bulk; and the right panel for all matter fields on the brane; for three different values of the compactification scales ($R^{-1}$ = 1 TeV (solid line), 2 TeV (dot-dashed line), and 10 TeV (dashed line)) as a function of the scale parameter {$t$}.}
\end{figure}
	
	
\par The evolution of the Yukawa couplings were derived in Refs.\cite{Abdalgabar:2013oja,Abdalgabar:2013xsa}, where the one-loop RGEs in the 2UED we study are: 
\begin{equation}
16 \pi^2 \frac{d Y_{i}}{d t} =  \pi \left( S(t)^2-1 \right) Y_{i} \left[ T_{i} - G_{i} + T \right],
\label{yukawa}
\end{equation}
where $i= u, d, e$ , $ T= 2(3 Tr ( Y_d^{\dagger} Y_d)+3 Tr ( Y_u^{\dagger} Y_u)+ Tr ( Y_e^{\dagger} Y_e))$ and the values of $G_i$ and $T_i$ are given in Tab.\ref{tab1}. 
\begin{table}
\caption{\label{tab1} \it The terms present in the various Yukawa evolution equations, see Eq.(\ref{yukawa}).}
\begin{center}
\begin{tabular}{cccccc}
\hline
 $Scenarios$ &  $G_u$ & $G_d$ & $G_e$ & $T_u=-T_d$ &$T_e$ \\
\hline
 $Bulk$ & $\frac{5}{6}g_1^2+\frac{3}{2}g_2^2+\frac{32}{3}g_3^2$ & $\frac{1}{30}g_1^2+\frac{3}{2}g_2^2+\frac{32}{3}g_3^2$ & $\frac{27}{30}g_1^2+\frac{3}{2}g_2^2$ & $3( Y_d^{\dagger} Y_d - Y_u^{\dagger} Y_u)$ & $3 Y_e^{\dagger} Y_e$ \\
\hline
 $Brane$ & $4( \frac{17}{20}g_1^2+\frac{9}{4}g_2^2+8g_3^2)$ & $4(\frac{1}{4}g_1^2+\frac{9}{4}g_2^2+8g_3^2)$ & $4(\frac{9}{4}g_1^2+\frac{9}{4}g_2^2)$ & $6( Y_d^{\dagger} Y_d - Y_u^{\dagger} Y_u)$ & $6 Y_e^{\dagger} Y_e$ \\
 \hline
\end{tabular}
\end{center}
\end{table}
 That is, when the energy scale $\mu > \frac{1}{R}$ or when the energy scale parameter $ t > \ln (\frac{1}{M_Z R})$, we shall use Eq.(\ref{yukawa}), however, when the energy scale $M_Z < \mu < \frac{1}{R}$, the Yukawa evolution equations are dictated by the usual SM ones, see Refs.\cite{ Abdalgabar:2013oja, Abdalgabar:2013xsa,  Cornell:2012qf}.

\par Yukawa coupling matrices can be diagonalised by using two unitary matrices {$U$} and {$V$}, where 

 $$
 UY^{\dagger}_u Y_u U^{\dagger}= \mathrm{diag}(f^2_u,f^2_c,f^2_t);\qquad VY^{\dagger}_d Y_d V^{\dagger}= \mathrm{diag}(h^2_d,h^2_s,h^2_b).
 $$
 \noindent
 The CKM matrix appears as a result (upon this diagonalisation of quark mass matrices) of $V_{CKM} = U V^{\dagger}$.
  The variation of the CKM matrix and its evolution equation for all matter fields in the bulk is \cite{Abdalgabar:2013laa, Babu:1987im}:

\begin{equation}
16 \pi^2 \frac{dV_{i \alpha}}{dt}= -6(\pi (S^2-1)+1) \left[\sum_{\beta, j \neq i}{ \frac{f_i^2+f_j^2}{f_i^2-f_j^2} h_{\beta}^2 V_{i\beta} V^\ast_{j \beta}}V_{j \alpha}+\sum_{j, \beta\neq \alpha}{\frac{h_{\alpha}^2+h_{\beta}^2}{h_{\alpha}^2-h_{\beta}^2} f_j^2 V^\ast_{j\beta} V_{j \alpha}V_{i \beta}}\right].
\label{ckm}
\end{equation}
 For all matter fields on the brane, the CKM evolution is the same as Eq.(\ref{ckm}) but multiplied by 2.

 The mixing matrix $V_{CKM}$ satisfies the unitarity condition, providing the following constraint

\begin{equation}
V_{ud} V^{\ast}_{ub} + V_{cd} V^{\ast}_{cb} + V_{td} V^{\ast}_{tb} = 0,
\end{equation}
that is, we have a triangle in the complex plane and the three inner angles $\alpha$, $ \beta$ and $\gamma$ are given by

\begin{equation}
\sin \beta = \frac{J}{|V_{td}| |V^{\ast}_{tb}| |V_{cd}| |V^{\ast}_{cb}|},\qquad \sin \gamma = \frac{J}{|V_{ud}| |V^{\ast}_{ub}| |V_{cd}| |V^{\ast}_{cb}|},
\end{equation}

\noindent with $\alpha = \pi - \beta - \gamma$. The shape of the unitarity triangle can be used as a tool to explore new
symmetries or other interesting properties that give a deeper insight into the physical content of new physics models.\\
\par On the other hand, in the quark sector both the mass ratios are related to mixing angles as
\begin{equation}
\theta_{13} \sim \frac{m_d}{m_b}, \,\qquad \theta_{23} \sim \frac{m_s}{m_b}.
\end{equation}
In Refs.\cite{Liu:2009vh, Liu:2011gra} a set of renormalisation invariants is constructed 
\begin{equation}
R_{13}= \sin(2\theta_{13}) \sinh \left[\ln \frac{m_b}{m_d}\right] \sim constant, \qquad R_{23}= \sin(2\theta_{23}) \sinh\left[\ln \frac{m_b}{m_s}\right] \sim constant.
\end{equation}

\section{Results}
 \par For our numerical calculations we set the compatification radii to be {$R^{-1}= 1$} TeV, 2 TeV and 10 TeV.
 Only some selected plots will be shown and we will comment on the other similar
cases not explicitly presented. We quantitatively  anlayse these quantities in 2UED models, though we observed similar behaviours for all values of {\textit{$R^{-1} $}}. The initial values we shall adopt at the $M_Z$ scale can be found in Ref.\cite{Xing:2007fb}.

\begin{figure}[h!]
\begin{center}
\includegraphics[width=7.5cm,angle=0]{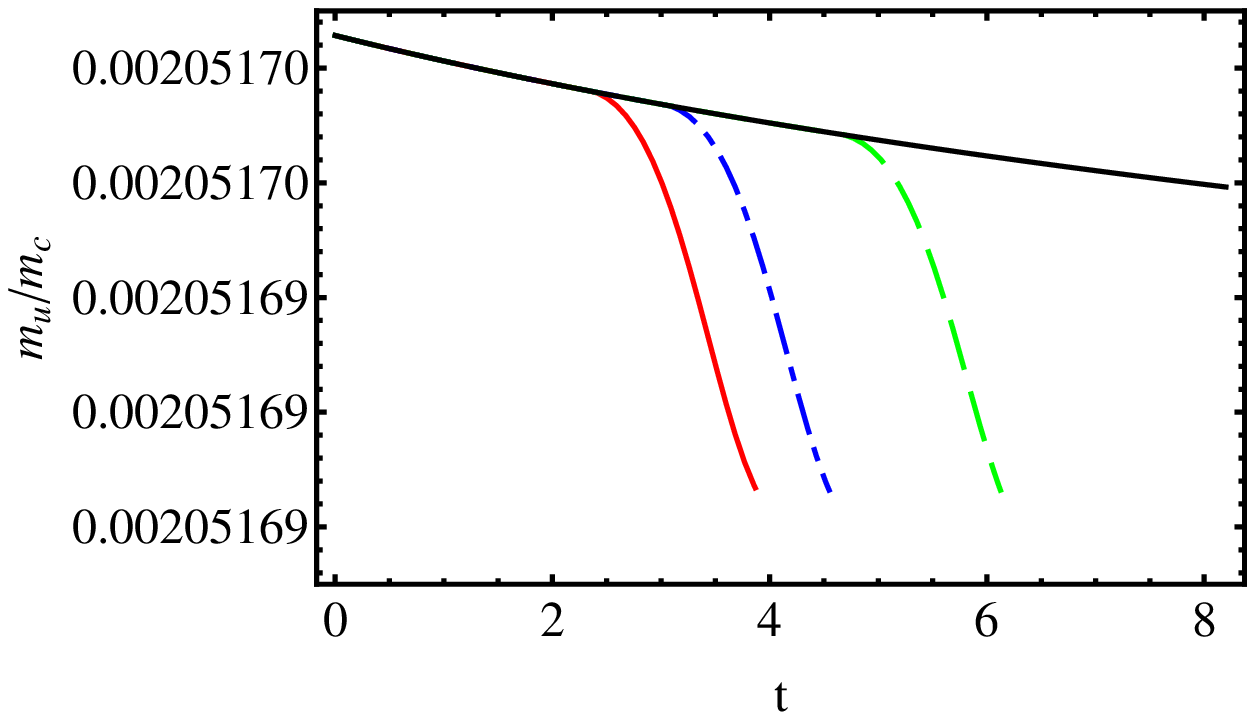} \qquad 
\includegraphics[width=7.5cm,angle=0]{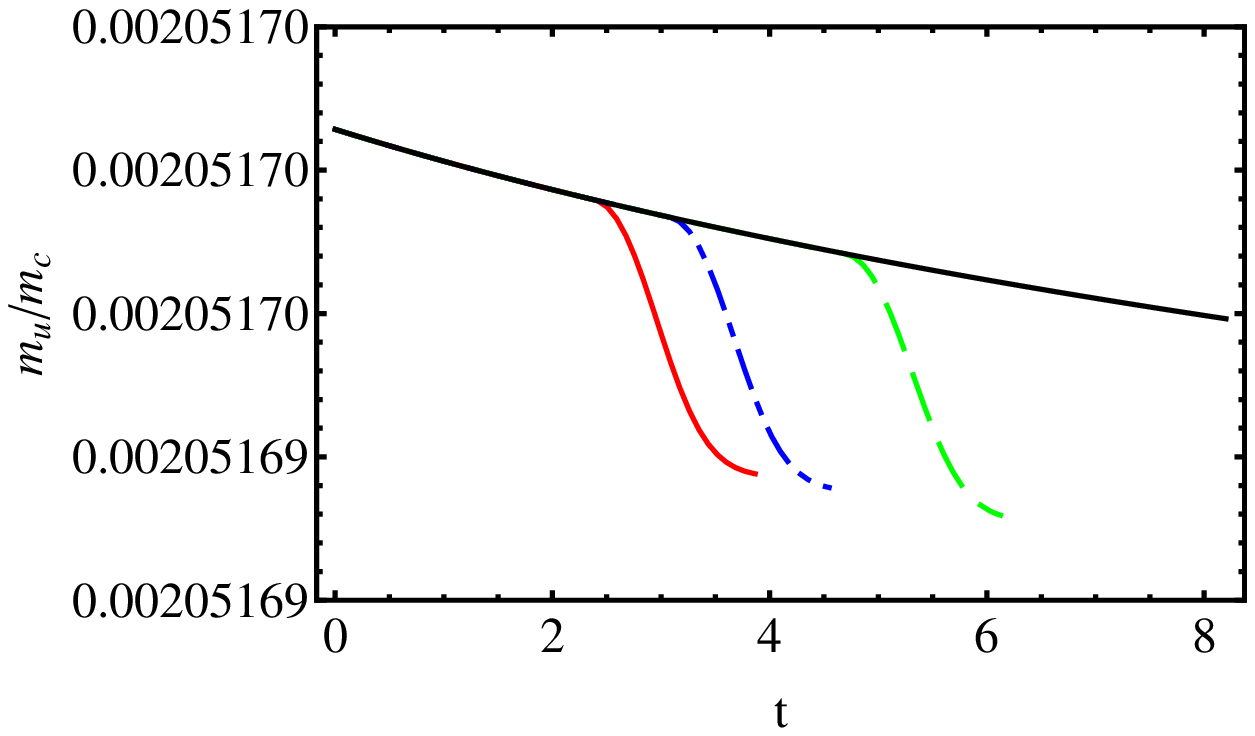}
\end{center}
\caption{\label{fig2} \it(Colour online) Evolution of the mass ratio {$\frac{m_u}{m_c}$} with: in the left panel all matter fields in the bulk; and the right panel for all matter fields on the brane. Three different values of the compactification radius have been used $R^{-1}$ = 1 TeV (solid line), 2 TeV (dot-dashed line), and 10 TeV (dashed line), all as a function of the scale parameter {$t$}.}
\end{figure}
\begin{figure}[h!]
\begin{center}
\includegraphics[width=7.5cm,angle=0]{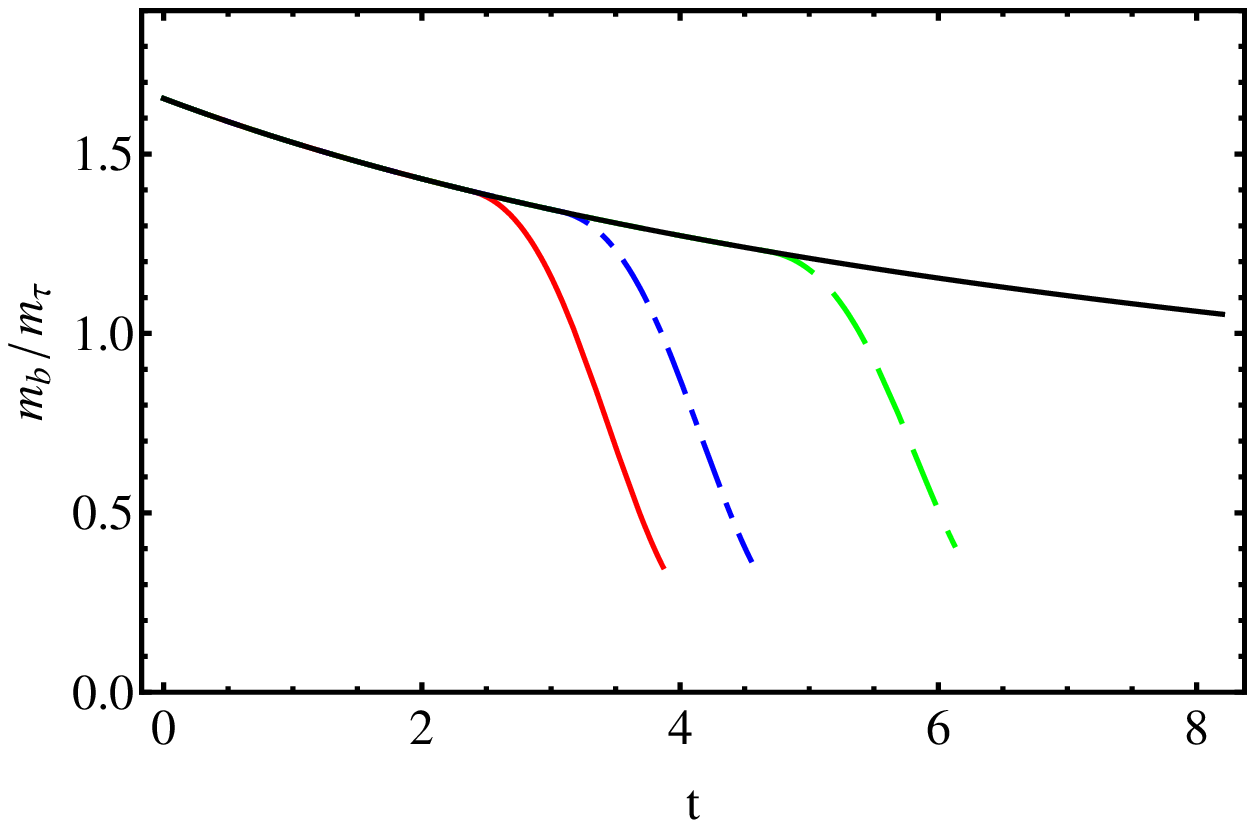} \qquad
\includegraphics[width=7.5cm,angle=0]{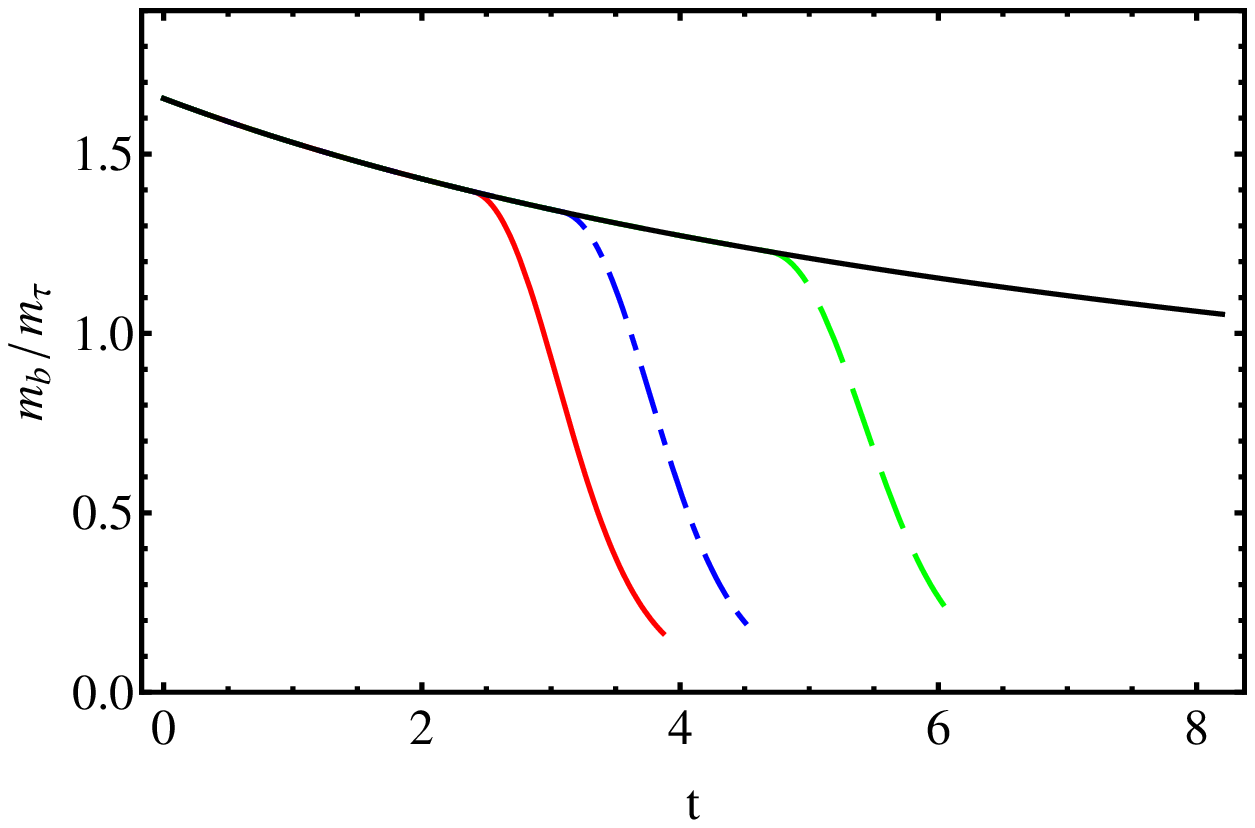}
\end{center}
\caption{\label{fig7}\it (Colour online) Evolution of the mass ratio {$\frac{m_b}{m_{\tau}}$}, with the same notations as Fig.2}
\end{figure}
\begin{figure}[h!]
\begin{center}
\includegraphics[width=7.5cm,angle=0]{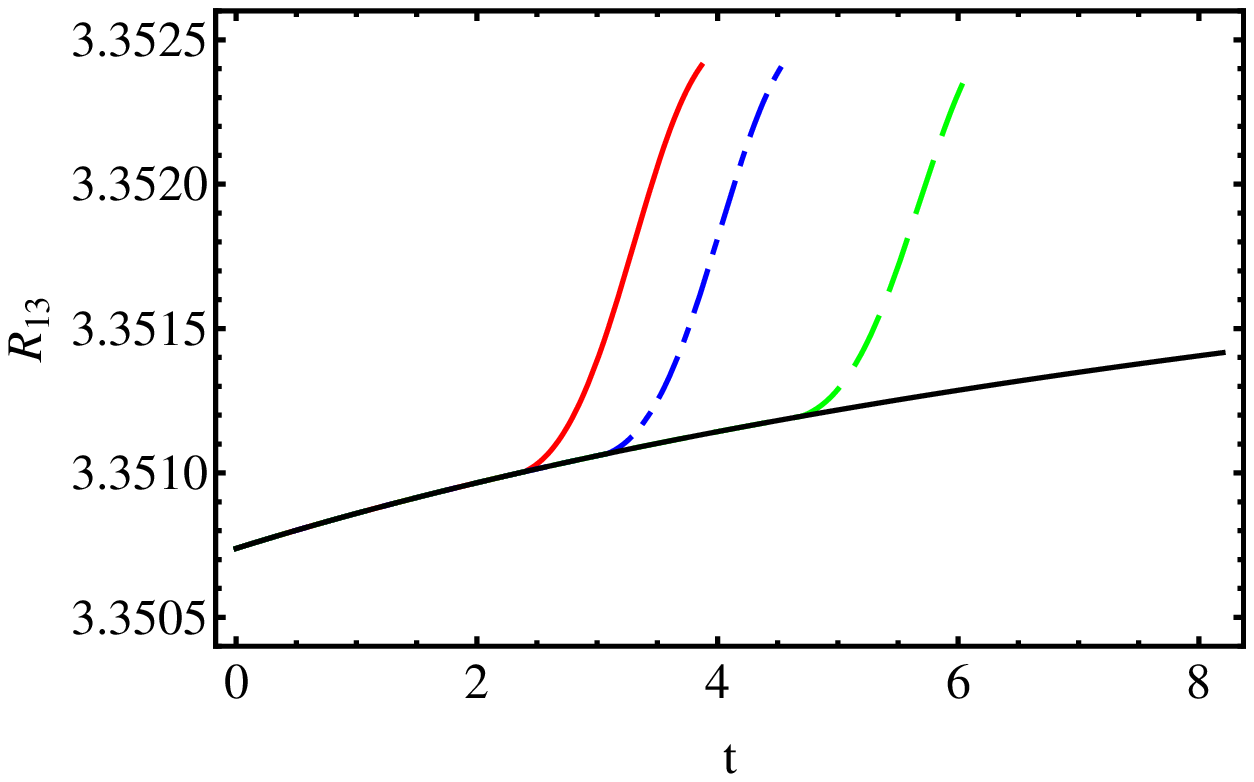} \qquad
\includegraphics[width=7.5cm,angle=0]{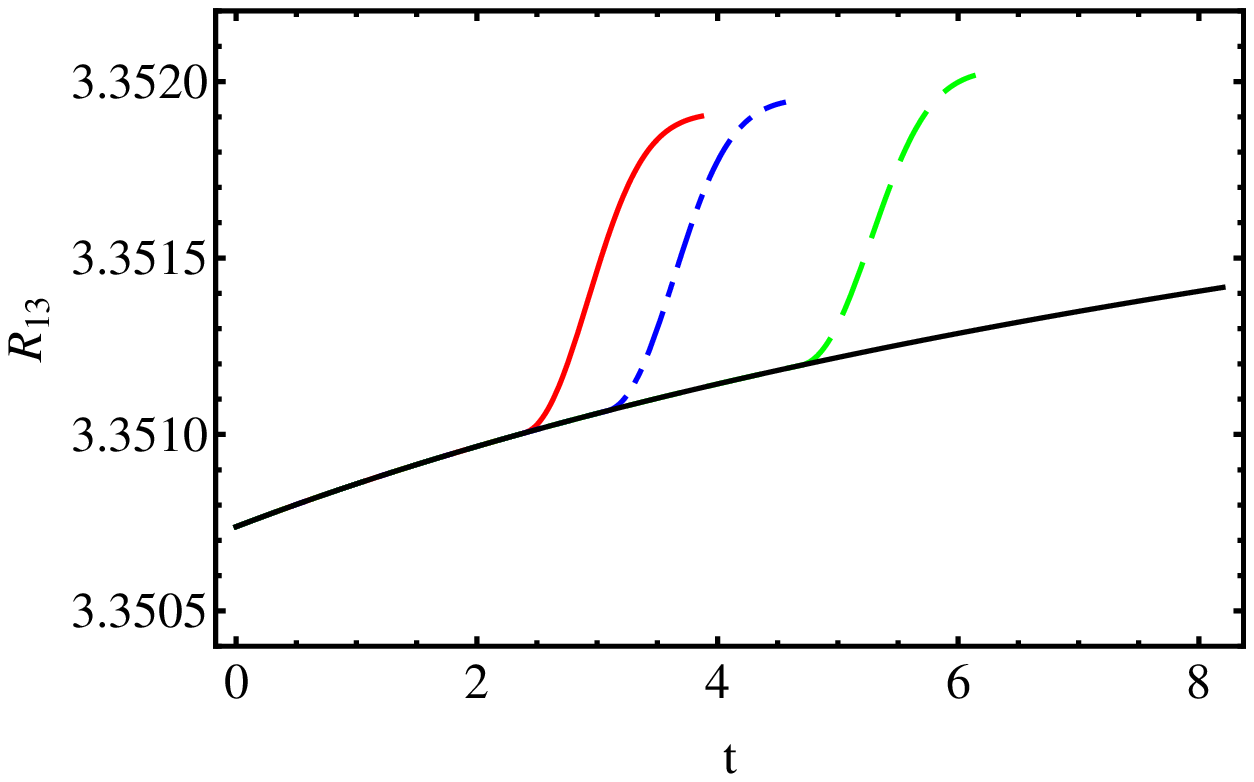}
\end{center}
\caption{\label{fig9}\it (Colour online) Evolution of the  {$R_{13}$}, with the same notations as Fig.2}
\end{figure}
\begin{figure}[h!]
\begin{center}
\includegraphics[width=7.5cm,angle=0]{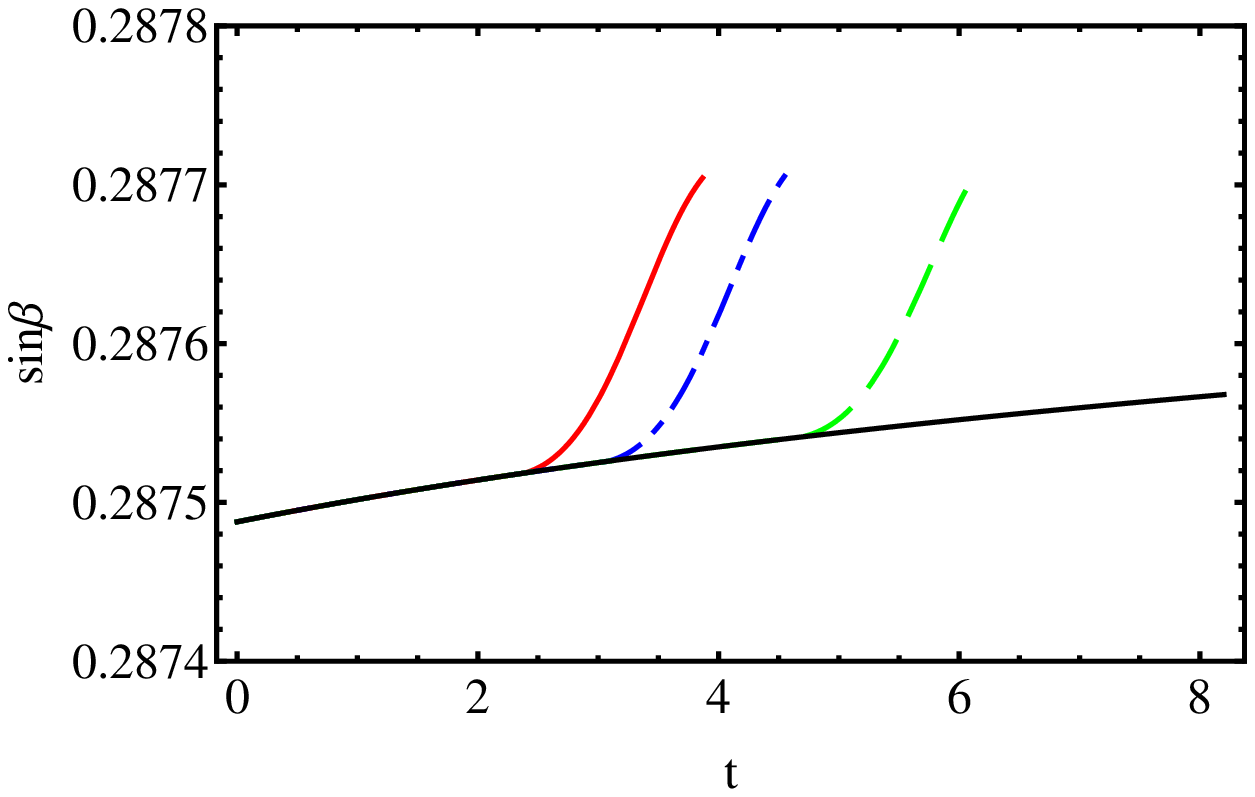} \qquad
\includegraphics[width=7.5cm,angle=0]{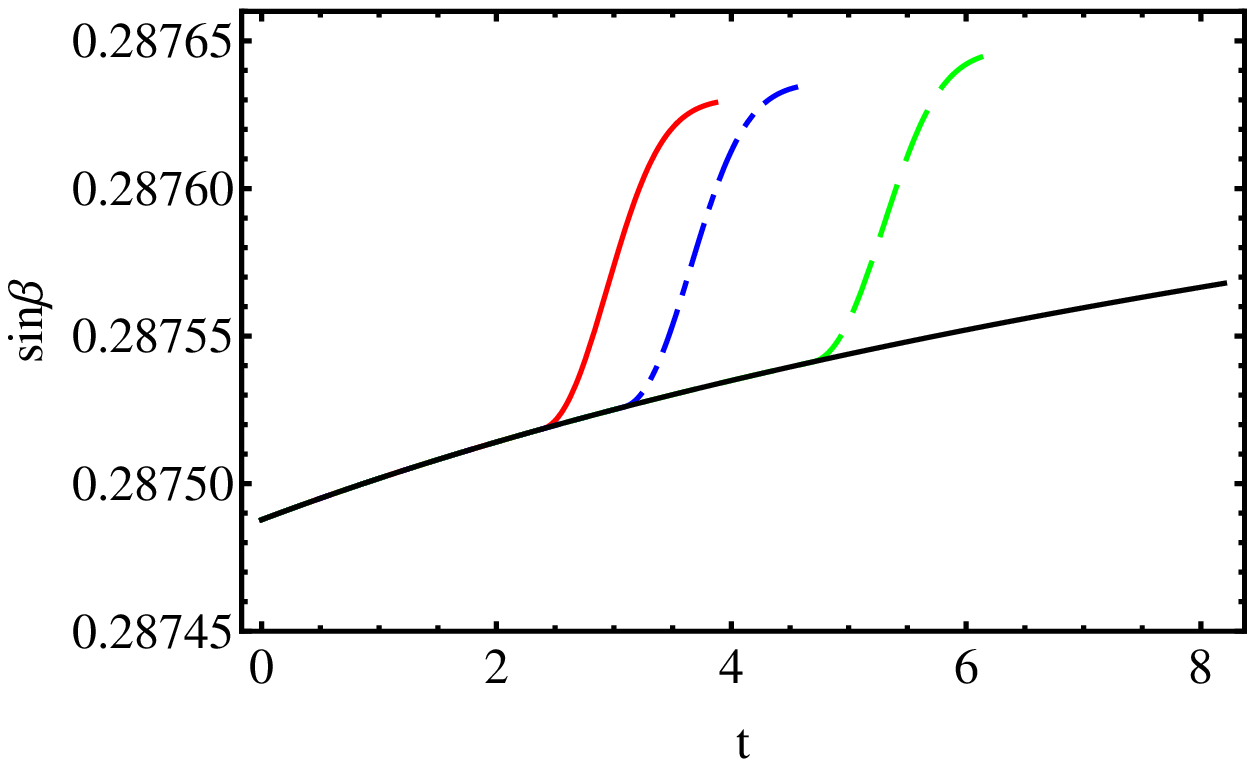}
\end{center}
\caption{\label{fig8}\it (Colour online) Evolution of {$\sin\beta$}, with the same notations as Fig.2}
\end{figure}


\section{Discussions and Conclusions}
\par As illustrated in Figs.\ref{fig1} \ref{fig2}, \ref{fig7}, the gauge couplings and  mass ratios evolve in the usual logarithmic fashion when the energies are below 1 TeV, 2 TeV and 10 TeV respectively. However, once the first KK threshold is reached the contributions from the KK states become increasingly significant and the effective 4D SM couplings begin to deviate from their normal trajectories. One finds that the running behaviours of the mass ratios are governed by the combination of the third family Yukawa couplings and the CKM matrix elements. This implies that the mass ratios of the first two light generations have a slowed evolution well before the unification scale. Beyond that point, their evolution  diverges due to the faster running of the gauge couplings, where any new physics would then come into play, and we find the scaling dependence of $\frac{m_d}{m_s}$ and  $\frac{m_e}{m_{\mu}}$ is very slow. 

\par On the other hand, Grand Unification Theories (such as $SU(5)$ and $SO(10)$) imply the well-known quark-lepton symmetric relation for fermion masses $m_d = m_e$. Due to power law running of the Yukawa couplings, the renormalisation effects on these relations can be large for {$\frac{m_b}{m_{\tau}}$}, for both scenarios, see Fig.\ref{fig7}. We have shown by numerical analysis of the one-loop calculation that the mass ratio {$\frac{m_b}{m_{\tau}}$}, as one crosses the KK threshold at {$\mu= R^{-1}$} for both scenarios, results in a rapid approach to a singularity before the unification scale is reached, which agrees with what is observed in the SM, however, the mass ratios decrease at a much faster rate. Note that we observed similar behaviour for {$\frac{m_d}{m_{e}}$} and {$\frac{m_s}{m_{\mu}}$}.

\par Let us now focus on the evolution of the set of renormalisation invariants $R_{13}$ and $R_{23}$ that
describe the correlation between the mixing angles and mass ratios to a good approximation. With a variation of the order of $\lambda^4$ and $\lambda^5$ under energy scaling respectively, as shown in Fig.\ref{fig9}, the energy scale dependence is weak because the increase of the mixing angles are compensated by the deviation of the mass ratios. Therefore the effect is not large.

\par In Fig.\ref{fig8} we present the evolution of the inner angle from the electroweak scale to the unification scale by using the one-loop RGE for the 2UED model, and demonstrate that the angle has a small variation against radiative corrections. To be more precise, the relative deviation for $\sin \beta$ is only up to  $0.05 \%$   in the whole range studied. Similar analysis can also be found for the angles $\alpha$ and $\gamma$. This result makes sense, since both the triangle's sides and area become larger and larger when the energy scale increases, the unitarity triangle (UT) is only rescaled and its shape does not change much during the RG evolution. The fact that inner angles are rather stable against radiative corrections indicates that it is not possible to construct an asymptotic model with some simple, special form of the CKM matrix from this simple scenario. The stability against radiative corrections suggests that the shape of the UT is almost unchanged from RGE effects. In this regards, the UT is not a sensitive test of this model in current and upcoming experiments.


\par In conclusion, the mass ratios in the 2UED model, with different possibilities for the matter fields, were discussed, where they are either bulk propagating or localised to the brane. We found that the 2UED model has substantial effects on the scaling of fermion masses for both cases, including both quark and lepton sectors. We quantitatively analysed these quantities for $R^{-1}$ = 1 TeV, 2 TeV and 10 TeV, observing similar behaviours for all values of the compactification radius. We have shown that the scale dependence is not logarithmic, it shows a power law behaviour. We also found that for both scenarios the theory is valid up to the unification scale, leading to significant RG corrections.

\section*{Acknowledgments}

We would like to thank our collaborators Aldo Deandrea and Ahmad Tarhini for the helpful discussions. This work is supported by the National Research Foundation (South Africa).


\end{document}